\documentstyle{ws-acs}

\begin{document}

\markboth{Pugnaloni, Barker and Mehta}{Multi-particle structures
in non-sequentially reorganized hard sphere deposits}

\catchline

\title{Multi-particle structures in non-sequentially reorganized hard sphere deposits}

\author{Luis A Pugnaloni}
\address{Instituto de Fisica de Liquidos y Sistemas Biologicos,\\
UNLP-CONICET, Casilla de correo 565, 1900 La Plata, Argentina}

\author{G C Barker}
\address{Institute of Food Research\\ Norwich Research Park,
Colney,\\
Norwich NR4 7UA, UK}

\author{Anita Mehta}
\address{S N Bose National Centre for Basic Sciences\\
Block JD Sector III Salt Lake,\\Calcutta 700 098, India}

\maketitle

\pub{Received (received date)}{Revised (revised date)}

\begin{abstract}
We have examined extended structures, bridges and arches, in
computer generated, non-sequentially stabilized, hard sphere
deposits. The bridges and arches have well defined distributions
of sizes and shapes. The distribution functions reflect the
contraints associated with hard particle packing and the details
of the restructuring process. A subpopulation of string-like
bridges has been identified. Bridges are fundamental
microstructural elements in real granular systems and their sizes
and shapes dominate considerations of structural properties and
flow instabilities such as jamming.
\end{abstract}

\section{Introduction}

There has always been a fascination, amongst physicists, with the
structures and configurations that exist within disordered
packings of hard particles (see for example, \cite{C+C}). One
interest stems from the fundamental, frustrated, geometries that
exist within sphere packings, e.g. \cite{Berryman},
\cite{Torquato}; another comes from the parallels between random
packings and the structures of real disordered materials like
liquids, glasses and granular solids, \cite{Bernal},
\cite{Seidler}. In particular it is clear that the mechanical and
transport properties of mesoscale disordered materials, like
powders and deposits, are strongly dependent on the relative
positions and connectivity of the constituent particles. A
striking example of this interplay follows when several particles
combine to form an 'arch' or 'bridge' near to the outlet of a
gravity flow container and cause the flow to stop. This blocking
phenomenon has an enormous impact on a wide range of technological
and industrial processes; there have been many attempts to
quantify the effect and to optimize operational parameters like
the outlet size and the internal granular flow pattern, e.g.
\cite{Nedderman}. However there is little information on the
statistical details of the particulate configurations that are the
underlying cause of the blocking.

In two dimensions arches and bridges can be observed throughout
dense, random, packings of hard disks, e.g. \cite{BR}, and they
appear to be ubiquitous elements of stable granular structures. In
a recent report To et al., \cite{Pak}, described experiments in
which a jamming arch of monosized disks was repeatedly formed, in
gravitational flow, across the outlet of a conical,
two-dimensional, hopper. These experiments indicated that the
jamming arches had configurations that were similar to those of
self-avoiding random walks. This statistical appreciation was used
to obtain predictions of crucial macroscopic parameters like the
jamming probability.

Below we give some details of bridge structures formed in models
of hard sphere deposits and explore the role of bridges in
three-dimensional disordered packings which are stable under
gravity. We do not find 'diffusive' bridge configurations but we
have identified a special, chain-like, subpopulation of bridges.

\begin{figure}
\centerline{\psfig{file=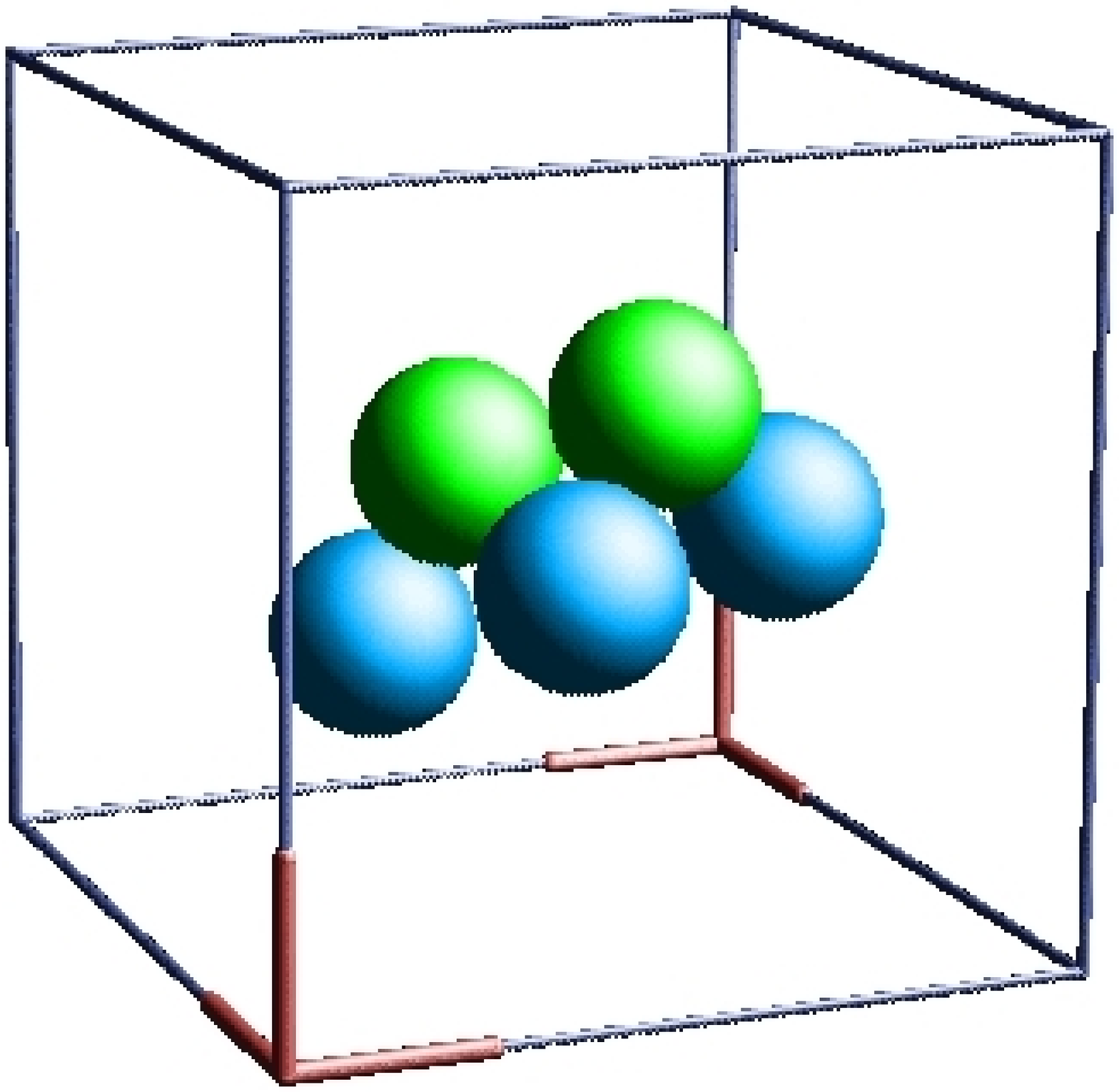,width=5.0cm,clip=}
\psfig{file=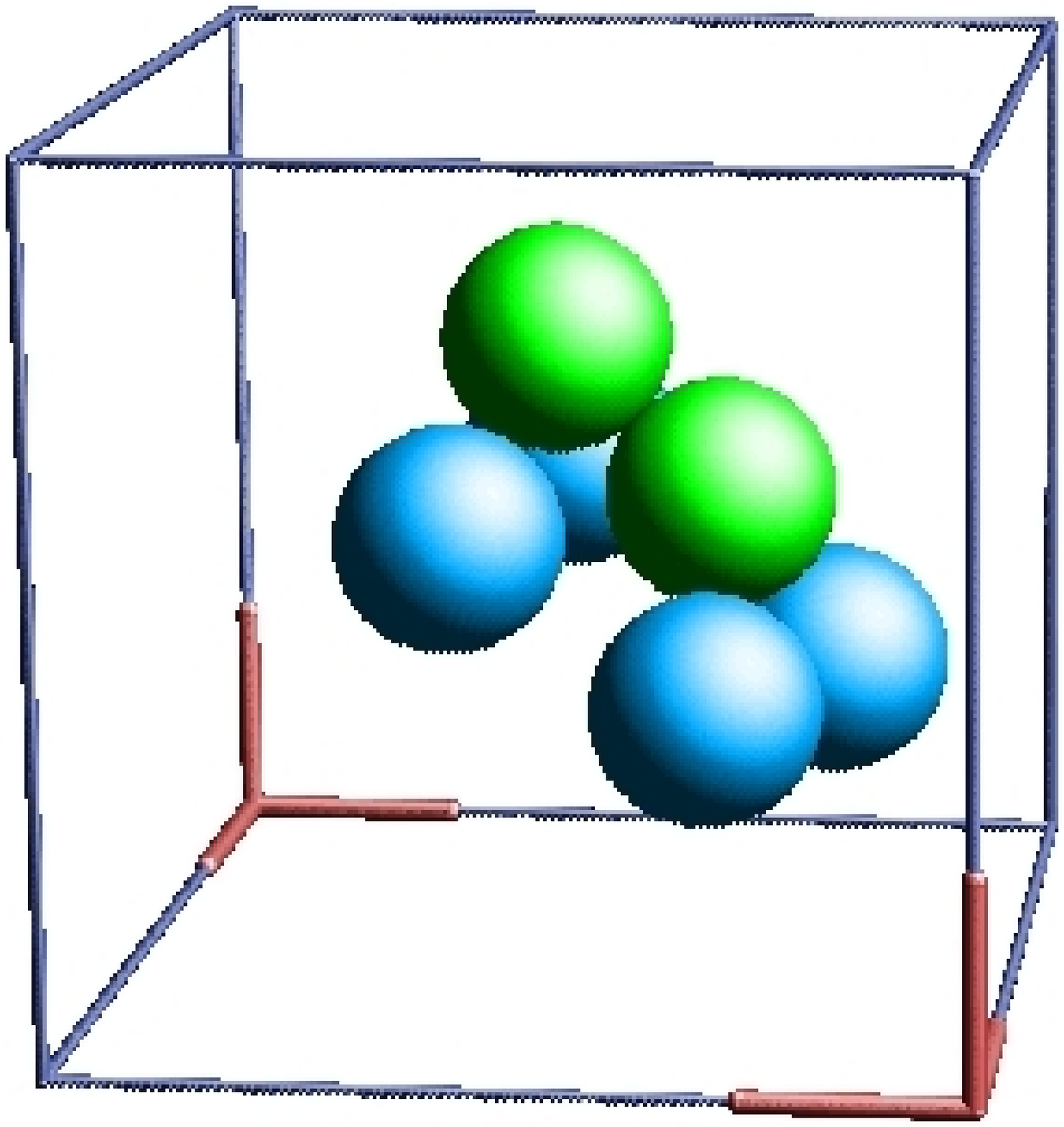,width=4.5cm,clip=}} \vspace*{8pt}
\caption{Simple two particle bridge with three and four base
particles} \label{fig1}
\end{figure}

In a stable packing of hard particles each particle rests on three
others in such a way that its weight vector passes through the
triangle formed from the three contact points (We do not consider
situations involving non-point contacts). A bridge is a
configuration in which the three point stability conditions of two
or more particles are linked i.e. there are mutual stabilizations.
In a simple example, with two particles A and B, particle A would
rest on particle B and two other particles whereas particle B
would rest on particle A and two further particles. Neither
particle A or B could rest, i.e. be part of the stable structure,
without the other. Bridge configurations, therefore, are the
result of non-sequential stabilizations; they cannot be formed by
the sequential placement of individual particles. In practice
almost all processing operations involving granular materials such
as pouring and shaking etc., are non-sequential processes. Two
examples of two particle bridges are shown in figure \ref{fig1}.
Each of these configurations is part of a large, dense, packing of
spheres; all those spheres not involved in the bridge have been
deleted so that it is clearly visible. The configuration on the
left uses only three particles as the base of the bridge whereas
that on the right uses four (base particles whilst ensuring the
stability of the configuration don't, themselves, involve mutual
stabilizations with other particles in the bridge structure). The
details of the bridge configurations, in terms of sizes and
shapes, are a manifestation of the volume and angular constraints
that exist in dense hard particle assemblies. In turn these
structures reflect the nature of the processing operations that
precede the formation of a stable packing. In this respect bridges
can be seen as part of the 'memory' of a granular system.

\section{Model deposits}
We have examined bridge structures in hard sphere assemblies that
are generated by an established, non-sequential, restructuring
algorithm, \cite{gcbam1,gcbam2}. This algorithm restructures a
stable, hard sphere, deposit in three distinct stages. Firstly
free volume is introduced homogeneously throughout the system and
the particles are given small, random, displacements. Secondly the
packing is compressed in a uniaxial external field using a
low-temperature Monte Carlo process. Thirdly the spheres are
stabilized using a steepest decent 'drop and roll' dynamics to
find a local minimum of the potential energy. Crucially, during
the third phase of the restructuring the spheres, although moved
in sequence, are able to roll in contact with spheres that are in
either stable or unstable positions; thus mutual stabilizations
may arise. The final configuration has a well defined network of
contacts and each sphere has a uniquely defined three point
stability (In practice the final configuration may include a few
'rattlers', \cite{Torquato}).

Restructuring simulations are performed in a rectangular cell (a
square prism) with periodic boundaries in the lateral directions
and a hard, disordered base perpendicular to the compression
(external field) direction. Our previous investigations,
\cite{gcbam1}, have shown that, this restructuring process does
not depend strongly on the simulation parameters and that, after
many cycles, restructured packings have a steady state described
by particular values for the structural descriptors such as the
mean packing fraction and the mean coordination number. Typically
the steady state mean volume fraction is in the range $\phi \sim
0.55 - 0.61$ and the mean coordination number is $Z \sim 5.6 \pm
0.1 $. The nature of the steady state is determined by the size of
the expansion phase or the 'amplitude' of the process,
\cite{gcbam1}. We have shown, \cite{gcbam2, gcbam3}, that the
random packings generated in this way have many features in common
with the states generated in vibrated granular media. In
particular we have shown that by varying the driving amplitude
systematically we can explore 'irreversible' and 'reversible'
branches of a density versus driving amplitude relationship
analogous to the experimentally observed behaviour, \cite{Nowak}.
We have used monosized spheres in order to avoid any problems with
induced size segregation and a disordered base prevents ordering.
In the packings we have considered here the $Q_{6}$ order
parameter, e.g. \cite{Torquato}, has a value
$Q_{6}/Q_{6}^{fcc}\sim0.05$. This small value indicates only a
very limited amount of face centered cubic crystalite formation in
the system ($Q_{6}^{fcc}$ is the value of the order parameter for
a face centered cubic crystal structure).

\section{Statistics of bridge structures}
We have identified clusters of mutually stabilized particles in
computer generated packings of hard spheres. Each configuration
includes approximately $N_{tot}=2500$ particles and we have
examined approximately $100$ configurations from each of two
steady states, with $\phi=0.56$ and $0.58$, of the reorganization
process.

\begin{figure}
\centerline{\psfig{file=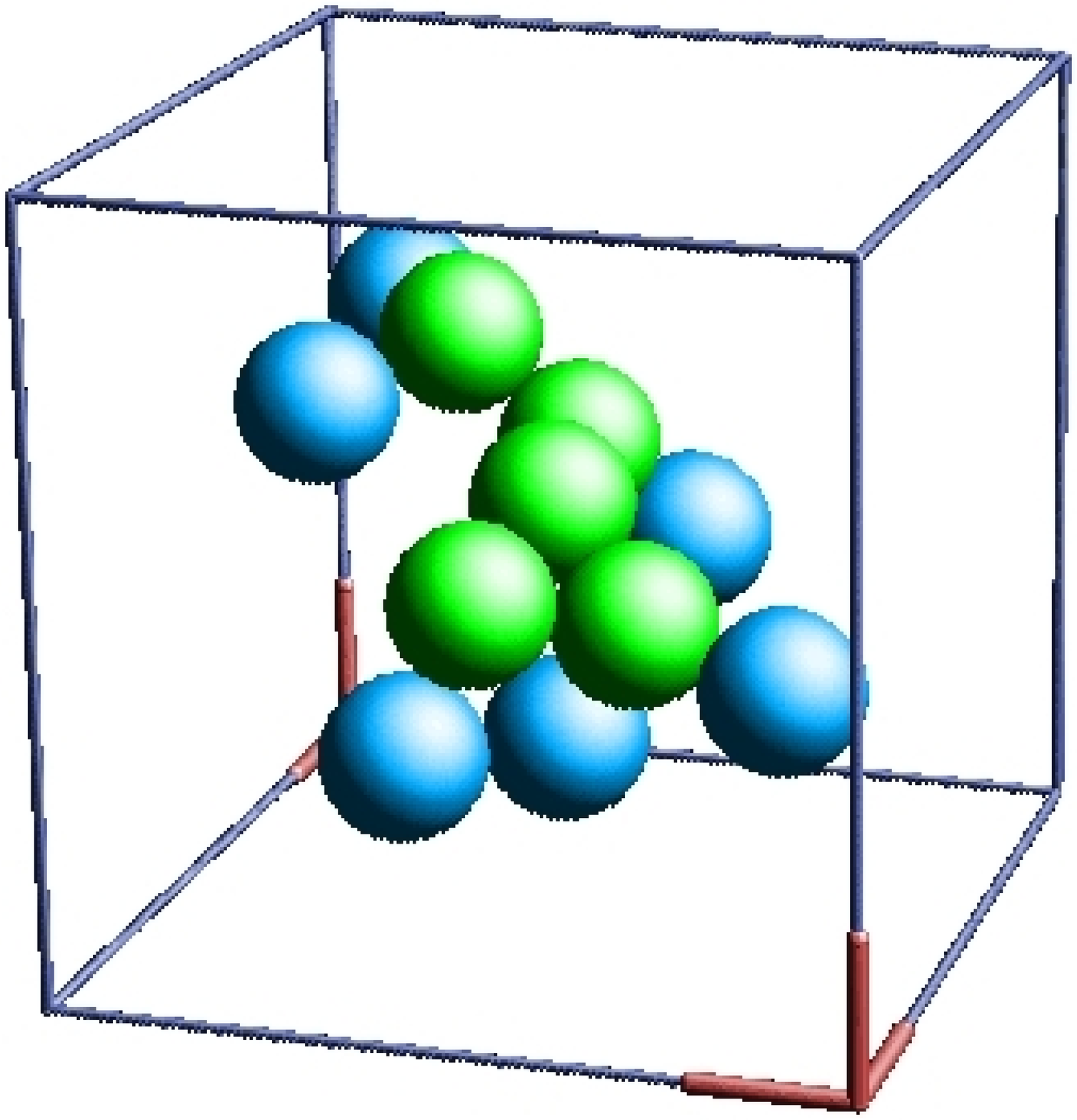,width=4.5cm,clip=}
\psfig{file=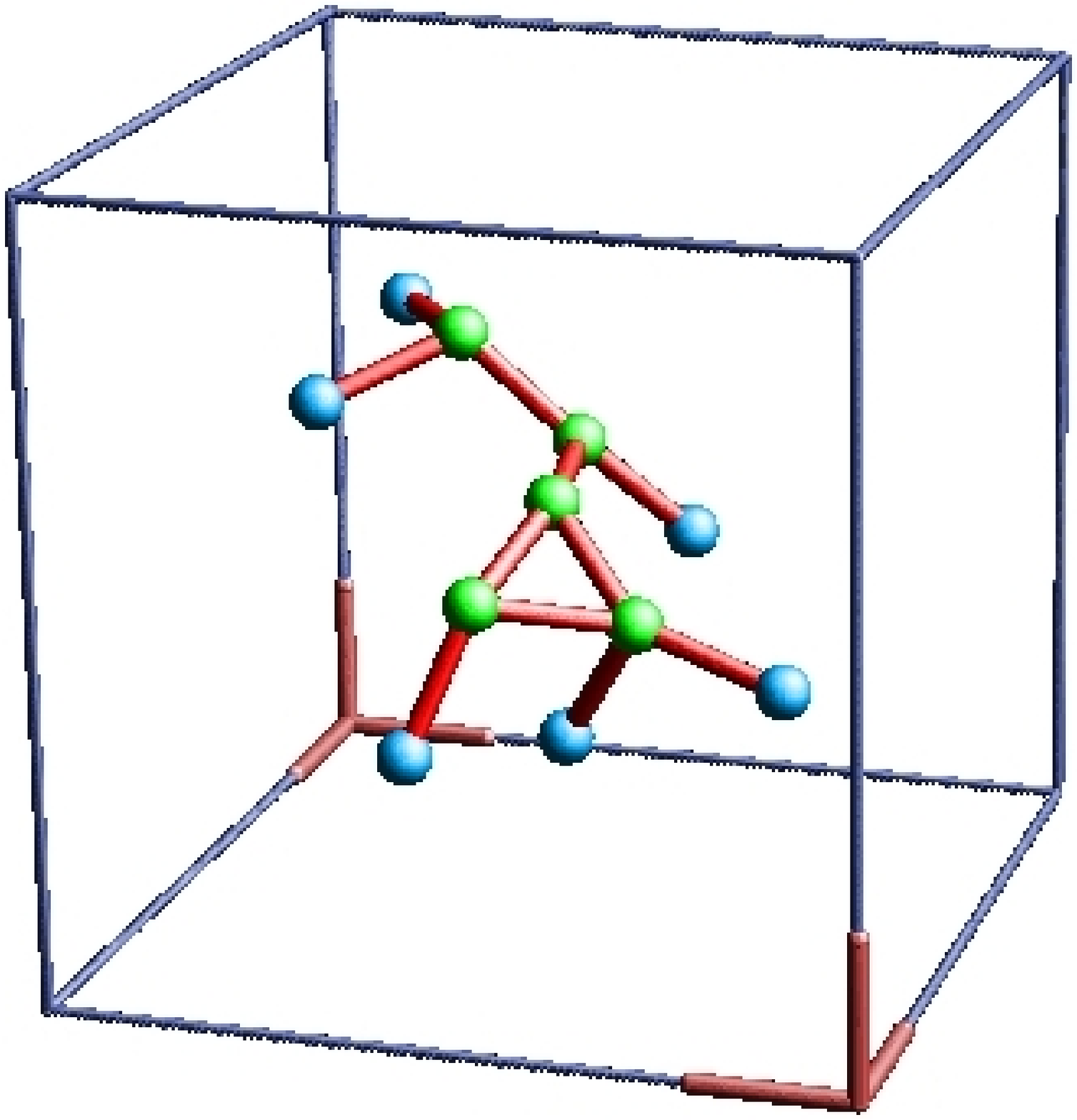,width=4.5cm,clip=}} \vspace*{8pt}
\caption{A five particle bridge with six base particles and the
corresponding contact network} \label{fig2}
\end{figure}

Figure \ref{fig2} illustrates a mutually stabilized cluster of
five particles that is part of a large, stable, packing; this
figure also shows six particles which form a base (all other
particles in the packing are hidden to make the diagram). Also
shown in figure \ref{fig2} is the network of contacts for the
particles in the bridge. This bridge is quite complex and
includes a set of three particles (lower and to the right) that
each have two mutual stabilizations.

\begin{figure}
\centerline{\psfig{file=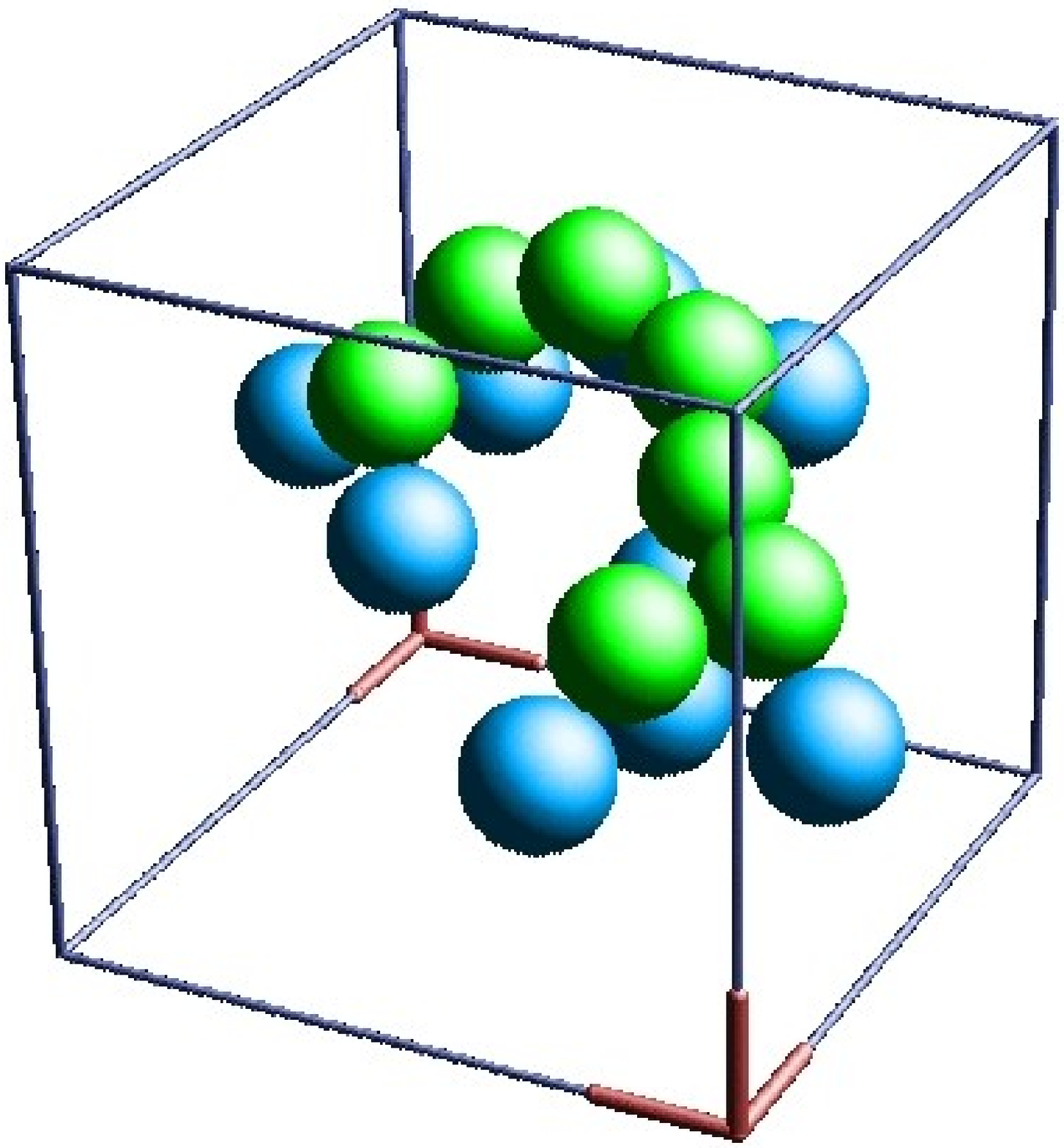,width=4.5cm,clip=}
\psfig{file=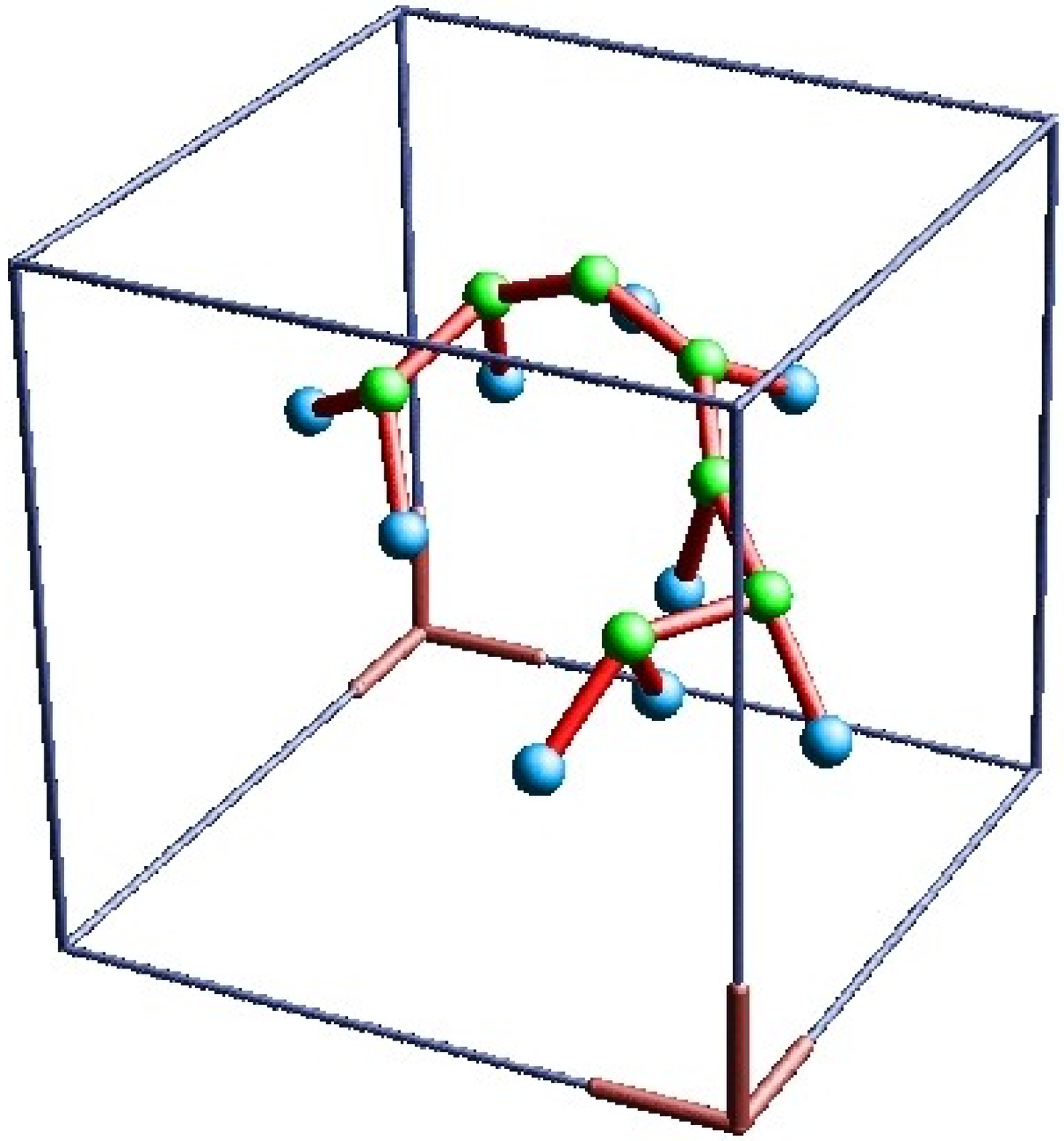,width=4.5cm,clip=}} \vspace*{8pt}
\caption{A seven particle string-like bridge with nine base
particles and the corresponding contact network} \label{fig3}
\end{figure}

Figure \ref{fig3} illustrates a seven particle bridge with nine
base particles. The contact network shows that although this
bridge is larger than that in figure \ref{fig2} it has a simpler
topology because all of the mutually stabilized particles are in
sequence - the bridge is string-like. The right hand configuration
in figure \ref{fig1}, with four base particles, is a string-like
bridge. In practice string like bridges are common; bridges such
as the one illustrated on the left hand side of figure \ref{fig1}
are very rare in our packings.

\begin{figure}
\centerline{\psfig{file=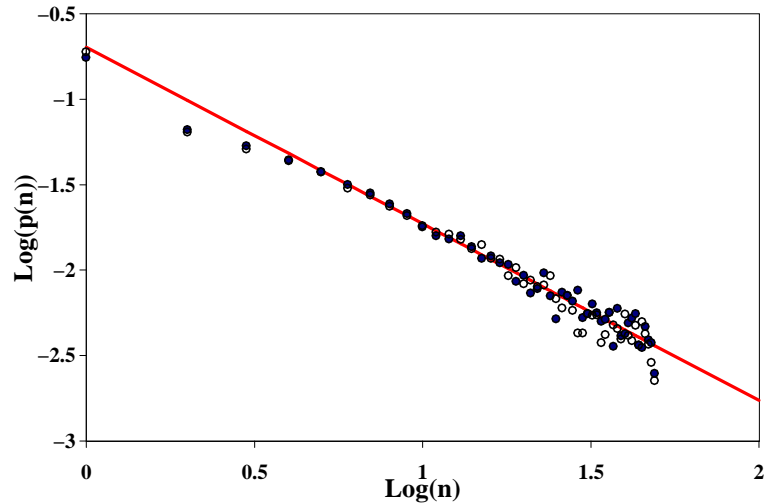,width=10cm}} \vspace*{8pt}
\caption{The size distribution of bridges in non-sequentially
reorganized hard sphere deposits; the full circles correspond to
packings with $<\phi>=0.58$, the open circles correspond to
packings with $<\phi>=0.56$ and the line is a fitted scaling
$p(n) \sim n^{-\alpha}$.} \label{fig4}
\end{figure}

Each packing contains a large variety of bridge sizes and shapes.
Approximately $80$ percent of particles are in mutually
stabilized locations. In figure \ref{fig4} we have plotted the
size distribution of the bridges as $Log(p(n))$ against $Log(n)$
where $p(n)=<nN(n)/N_{tot}>$ and $N(n)$ is the number of bridges
which contain $n$ mutually stabilized particles. Angular brackets
indicate an average over configurations in the steady state. We
can consider $p(n)$ as the probability that a particular particle
is included in a bridge with size $n$. Over a wide range of bridge
sizes the distribution function has a scaling behaviour of the
form $p(n) \sim n^{-\alpha}$ with $\alpha \sim 1.0 \pm 0.03$. The
bridge size distribution is not strongly dependent on the volume
fraction of the packings.

\begin{figure}
\centerline{\psfig{file=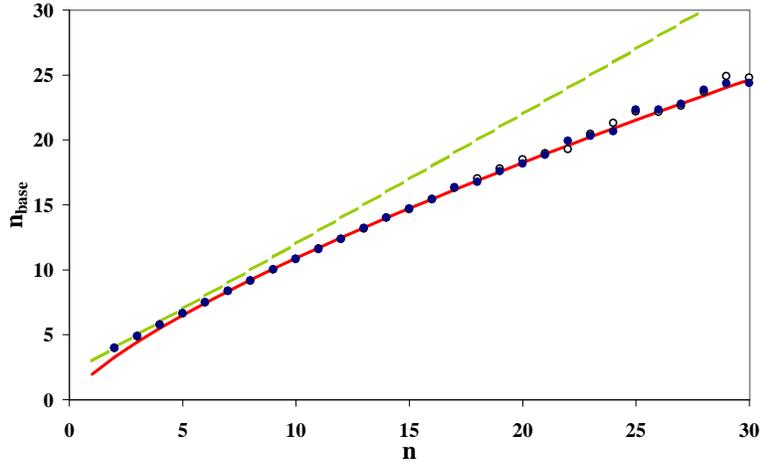,width=10cm,clip=}}
\vspace*{8pt} \caption{The mean base size for bridges with size
$n$;the dashed line indicates behaviour for string-like bridges
$n_{base} = n+2$ and the full line is a scaling fitted to the
behaviour of the larger bridges.} \label{fig5}
\end{figure}

For a particular bridge size the number of base particles, which
complete the stabilization, is variable with an upper bound,
$n+2$, corresponding to a string-like bridge. The mean number of
base particles, $n_{base}$, is plotted as a function of the bridge
size in figure \ref{fig5}. There is a crossover in behaviour at $n
\sim 8$; small bridges are predominantly string-like and larger
bridges have more complex structures with relatively fewer base
particles. Again this property is not strongly dependent on the
volume fraction of the packings in the range we have considered.
We did not observe any 'domes' or 'canopies' although this could
be an artefact of the relatively small sizes of the deposits. For
a particular bridge configuration a triangulation of the base
particles can be used to construct a unique bridge axis as the
mean of the triangle normals. With respect to this axis
geometrical descriptors, such as the radius of gyration or the
aspect ratio, also show a cross over that indicates the
significance of a sub-population of string-like bridges.

\begin{figure}
\centerline{\psfig{file=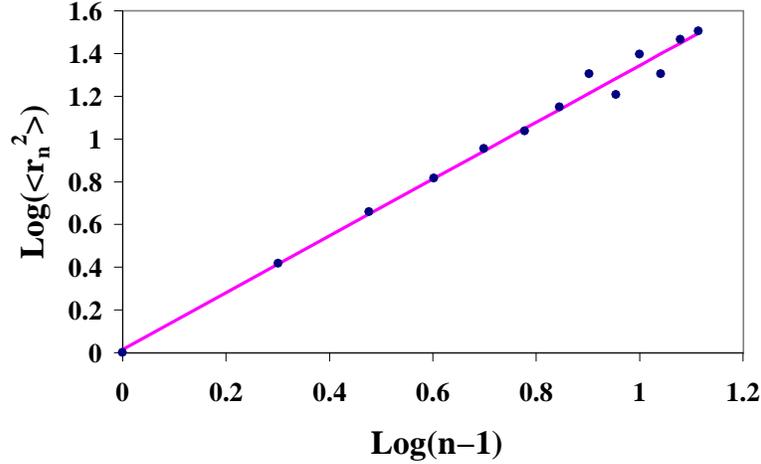,width=10cm,clip=}}
\vspace*{8pt} \caption{The mean squared displacement,
$<r_{n}^{2}>$, for string-like bridges as a function of the
number of mutually stabilizing bonds $n-1$. Bridges are part of
restructured deposits with steady state volume fraction
$\phi=0.58$.} \label{fig6}
\end{figure}

A string-like bridge has uniquely defined end particles and,
therefore, a well defined extension. The mean squared separation,
$<r_{n}^{2}>$, of the end particles for string-like bridges scales
with the number of stabilizing bonds according to $<r_{n}^{2}>
\sim (n-1)^\gamma$ with $\gamma = 1.33$. This 'superdiffusive'
behaviour is illustrated in figure \ref{fig6}. The population of
string-like bridges we observe, in reorganized three dimensional
deposits, is thus distinct from the random walk structures that
have been identified as the cause of blocking at the outlet of a
two dimensional hopper, \cite{Pak}.

\begin{figure}
\centerline{\psfig{file=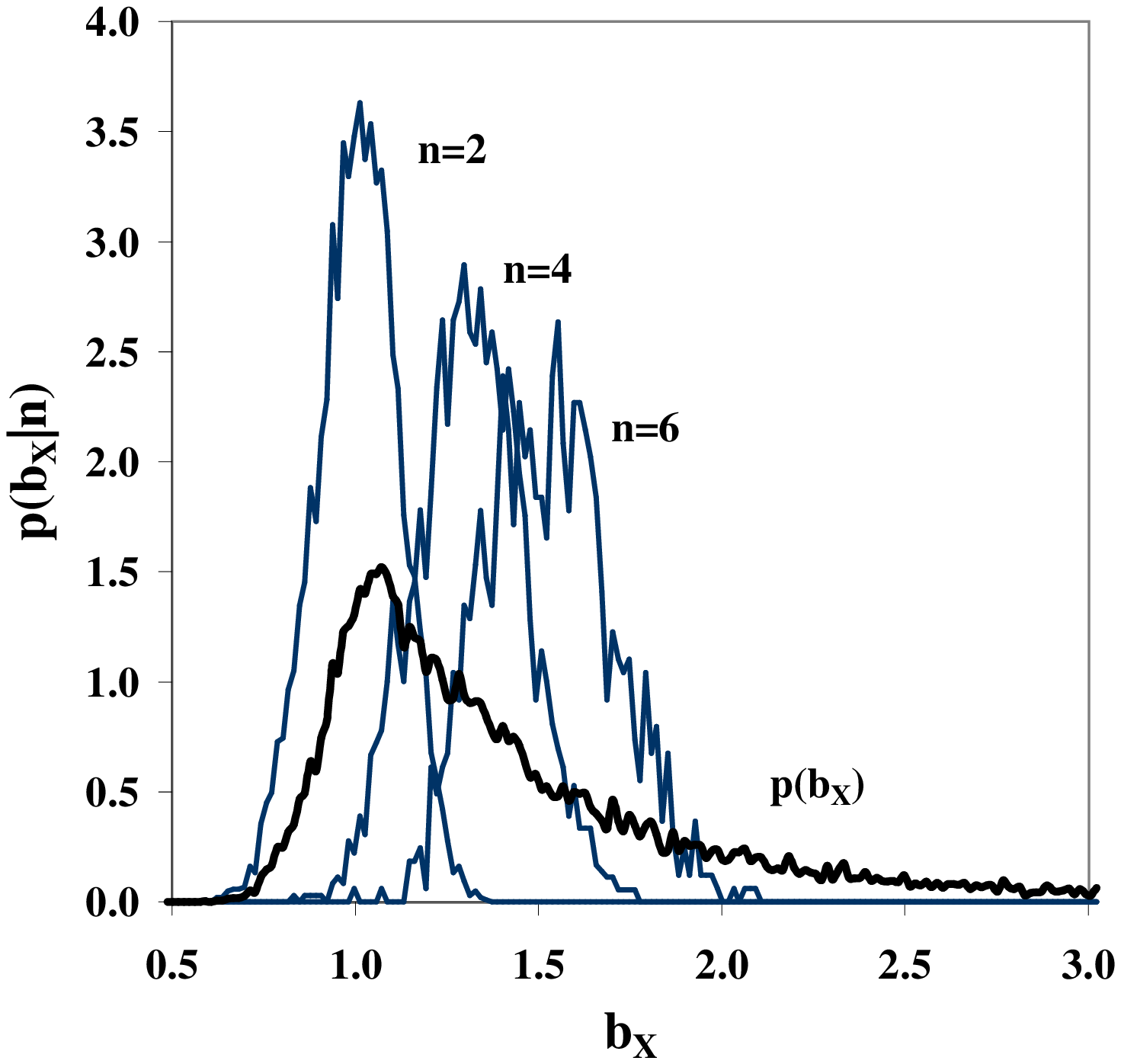,width=6.0cm,clip=}
\psfig{file=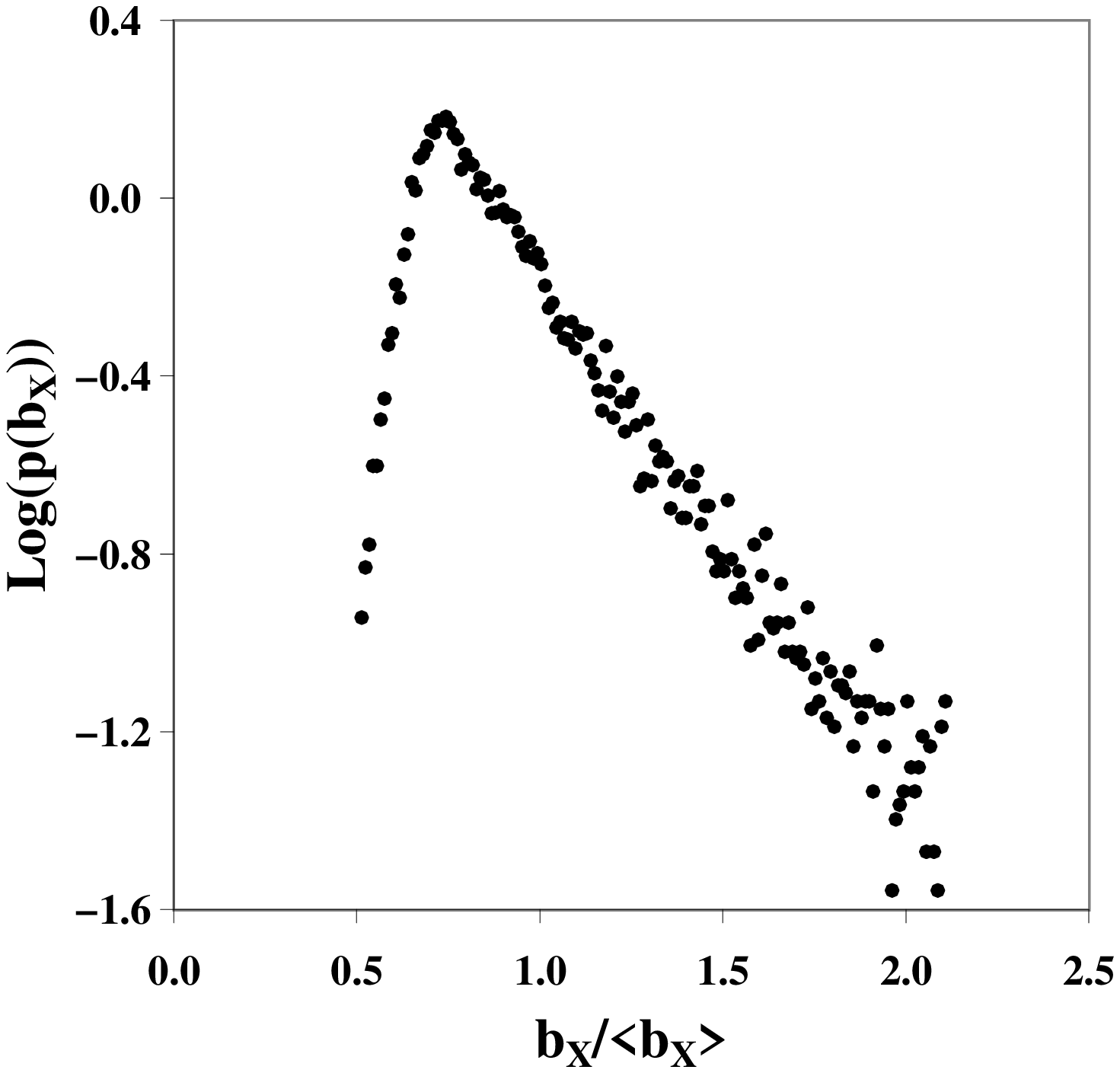,width=6.0cm,clip=}} \vspace*{8pt}
\caption{The distribution of base extensions for bridges that are
part of restructured deposits with steady state volume fraction
$\phi=0.58$. The left hand figure also shows the distributions
conditional on the bridge size, $n$, for $n=2,4,6$. The right
hand figure shows the logarithm of the density as a function of
the normalized variable $b_{x}/<b_{x}>$.} \label{fig7}
\end{figure}

Figure \ref{fig7} shows the distribution function of base
extensions for all bridges in packings that are part of the
restructuring steady state with $\phi=0.58$. The extension,
$b_{x}$, is the projection, in a plane perpendicular to the
external field, of the radius of gyration of the base particle
configuration (about the bridge axis). Clearly this measure is
related to the ability of a bridge to span an opening and,
therefore, is an indicator of the jamming potential for a bridge.
We have also shown, in figure \ref{fig7}, distribution functions
that are conditional on the bridge size $n$ (for $n=2,4,6$). The
conditional distributions are sharply peaked, and are bounded at
finite $b_{x}$, but the total distribution has a long tail, at
large extensions, reflecting the existence of large bridges. In
the second part of figure \ref{fig7} we have plotted the logarithm
of the probability density against a normalized variable,
$b_{x}/<b_{x}>$, where $<b_{x}>$ is the mean extension of bridge
bases. This figure emphasizes the exponential tail of the
distribution function and also shows that bridges with small base
extensions are unfavoured. The absence of bridges with base
extensions that are considerably smaller than the mean extension
is a reflection of the angular constraints that exist in hard
particle structures. Small base extensions reduce the number of
possible stable configurations for bridges with fixed size $n$.
The form of the distribution in figure \ref{fig7} can be
interpretted, clearly, in terms of a partition, $p(b_{x})=\sum_n
p(b_{x} \mid n)p(n)$, since the conditional probabilities have
restricted ranges, reflecting hard particle volume and angular
constraints, and the size distribution has a well defined scaling
that reflects the particular bridge creation and anhiliation
processes that are included in the restructuring. In this form it
is clear that the tail of the distribution of $b_{x}$ arises from
the summation and not from bridges with a particular size. It is
interesting to note that the form of the normalized distribution
in figure \ref{fig7} is similar to the distribution of the normal
forces in dense packings of hard particles, e.g. \cite{OHern}.

\section{Discussion}
Bridges and arches are significant elements of the mesostructure
in many granular solids processing scenarios, e.g.
\cite{BR,Cooper}. These structures, which extend beyond the scale
of single particles, are strongly associated with important
macroscopic properties of materials and with flow instabilities.
We have shown that bridge structures are included throughout the
non-sequentially reorganized deposits we have constructed. Bridges
have well defined statistics and, to a first approximation, they
are distributed homogeneously within the deposits. We have
identified a sub-population of bridges, which have string-like
configurations, that dominate for low bridge sizes. At present it
is unclear whether these structures are a property of the
particular reorganization scheme considered here or whether they
are a fundamental feature of non-sequential reorganization in hard
sphere deposition. The bridge size statistics we have presented do
not depend strongly on the volume fraction of the deposits but
other measures, such as the bridge orientations (which we will
report elsewhere), do vary with packing density i.e. with the
expansion amplitude of the reorganization process. Additionally it
is clear that non-sequential structures like bridges, that become
trapped in the close packed systems, frustrate local ordering in
packings of monosized spheres. Thus the onset of ordering must
coincide with changes in the distribution of bridges; for driving
amplitudes that are smaller than those used to construct the
deposits considered above we have observed the sudden onset of
ordering, \cite{gcbam4}. We have not examined correlations of the
bridges in the series of packings generated by the reorganization
process.

Clearly, based on an assumption that the bridges in bulk are the
same as those close to an opening, the statistics of extended
structures in hard particle deposits is sufficient to estimate the
probability that a bridge will form a span at an outlet of a fixed
size. In three dimensions this probability is not the same as the
probability that a bridge will form a blockage or a 'jam'. However
initial investigations, \cite{Luis}, indicate that data, analogous
to the complement of the cumulative form of the distribution in
figure \ref{fig7}, are in qualitative agreement with observations
of the jamming probability. We hope to present details of these
analyses in a future report.

\section*{Acknowledgements}
GCB acknowledges support from the Fundaci\'{o}n Antorchas, and the
hospitality of Prof. J. Raul. Grigera, during a visit to IFLYSIB,
La Plata, Argentina where some of this work was completed. LAP
acknowledges support from the International Union of Pure and
Applied Biophysics during a visit to IFR, UK.

\end{document}